\documentstyle[preprint,aps,epsf]{revtex}
\begin{document}
\setcounter{page}{0}

\draft

\title{Production of $\eta$ Mesons in Double Pomeron Exchange}
\author{Alec J. Schramm \thanks{Email: {\it alec@phys.oxy.edu}}}
\author{and}
\author{Daniel H. Reeves}
\address{Department of Physics, Occidental College, Los Angeles, CA 90041}


\maketitle
\thispagestyle{empty}

\begin{abstract} 

We estimate the production cross section for $\eta_c$ and $\eta_b$ mesons via pomeron-pomeron fusion in peripheral heavy-ion collisions.  Total and elastic ${\cal PP}$ cross sections are calculated in an equivalent pomeron approximation. 
\end{abstract}

In a recent paper \cite{oxy}, we discussed the production of the
pseudoscalar mesons $\eta_c$ and $\eta_b$ in peripheral collisions of high-energy heavy ions.  Applying an equivalent photon approximation to the large virtual photon luminosities accompanying the colliding nuclei, we calculated production cross sections for $\eta_c$ and $\eta_b$ via $\gamma\gamma$ fusion.  For the large center-of-mass beam energies available at the Relativistic Heavy Ion Collider (RHIC) at Brookhaven (100 GeV/nucleon) and the Large Hadron Collider (LHC) at CERN (3.5 TeV/nucleon), we found the $\gamma\gamma$ elastic cross sections for the $\eta_c$ to be about 0.8 mb at the LHC and 9 $\mu$b at RHIC; the cross sections for the heavier $\eta_b$ were on the order of $0.6 \mu$b and 0.2 nb, respectively.

There are backgrounds to this process which need to be considered, such as strong double diffractive scattering, i.e., double Pomeron exchange. These competing contributions are the subject of this work.  In particular, we use the effective pomeron approximation developed earlier \cite{us} to calculate the pomeron-pomeron cross sections for $\eta_c$ and $\eta_b$ production.  Of course, the detection of such processes requires the low background attendant in peripheral ion collisions; to this end, we calculate both the impact parameter dependence and the effects of nuclear absorption, complementing work of Natale \cite{natale}.

The equivalent pomeron approximation \cite{us} is a generalization of the standard equivalent photon approximation \cite{fermi}, in which one replaces the electromagnetic field of a relativistic charged particle with an equivalent pulse of real photons.  For a pomeron, one begins with the Regge trajectory
\begin{eqnarray}
\alpha_{_{\cal P}}(t) = 1 + \epsilon + \alpha_{_{\cal P}}^\prime t,
\end{eqnarray}
where $\epsilon = 0.085$ and $\alpha_{_{\cal P}}^\prime = 0.25$ GeV$^{-2}$. 
Note that this trajectory has an intercept close to 1, suggesting that the pomeron behaves like a spin-one boson in processes in which the exchanged four-momentum squared $k^2=t$ is small.  Pomeron exchange can thus be described to be like that of an isoscalar ``photon'', with propagator at high c.m. energy given by \cite{DL1},
\begin{eqnarray}
|D_{_{\cal P}}(t=-\vec{k}^2;s)| = (s/m^2)^\epsilon~e^{-r_0^2\vec{k}^2}
\end{eqnarray}
where $r_0$ is the pomeron range parameter \cite{us},
\begin{eqnarray}
r_0^2 = \alpha_{_{\cal P}}^\prime\ln{(s/m^2)}~.
\end{eqnarray}
Following Donnachie and Landshoff \cite{DL2}, we denote the nucleon-pomeron coupling by
\begin{eqnarray}
\beta_{NP} = 3\beta_0F_N(-t) ~,
\end{eqnarray}
where the quark-pomeron coupling $\beta_0$ is
\begin{eqnarray}
\beta_0 = 1.8 \mbox{GeV}^{-1} 
\end{eqnarray}
and $F_N(-t)$ is the isoscalar magnetic nucleon form factor.  
For nucleus-pomeron interaction, we appeal to the additivity of the total nucleon-nucleon cross sections and replace $\beta_0$ by $A\beta_0$ and $F_N(-t)$ by the elastic nuclear form factor $F(\vec{k}^2)$. 
We consider $^{208}$Pb nuclei, and approximate the form factor by the Gaussian \cite{dez}
\begin{eqnarray}
F(\vec{k}^2) = e^{-\vec{k}^2/2Q_0^2}~,
\end{eqnarray}
where $Q_0\approx 60$ MeV.   

In the spirit of the equivalent photon approximation, we write the cross section for ${\cal PP}$ fusion arising from the interaction of ion beams $A$ and $B$ with center of mass energy $s$ in the form (see Figure 1)
\begin{eqnarray}
\sigma^{\cal PP}_{AB} = 
\int~dx_1dx_2~f_{_{\cal P}}^A(x_1)f_{_{\cal P}}^B(x_2)\sigma_{_{\cal PP}}^X(s_{_{\cal PP}}) ~,
\end{eqnarray}
where $f_{_{\cal P}}(x)$ is the distribution function for finding a pomeron in the nucleus with energy fraction $x$.  
This was worked out in reference \cite{us}; for a nucleus of mass $M$ and nucleon number $A$,
\begin{eqnarray}
f_{\cal P}^A(x) = \left(\frac{3A\beta_0Q_0^2}{2\pi}\right)^2
                     \left(\frac{s^\prime}{m^2}\right)^{2\epsilon}
                     \frac{1}{x}~e^{-x^2M^2/Q_0^2} ~,
\end{eqnarray}
where $s^\prime$ denotes the invariant subprocess with which the pomeron participates.  
In this expression, $\sigma_{\cal PP}^X$  is the production cross section for the subprocess ${\cal PP}\rightarrow X$ with squared center of mass energy $s_{_{\cal PP}} \equiv x_1x_2s$ \cite{brodsky},
\begin{eqnarray}
\sigma_{_{\cal PP}}^{X} = \frac{8\pi^2}{M_X}~
      \Gamma_{_{X^0\rightarrow\cal PP}}~\delta(s_{_{\cal PP}} - M_X^2)~.
\end{eqnarray}
Relying on the ``pomeron as isoscalar photon'' identification, the partial two-pomeron decay width $\Gamma_{_{X^0\rightarrow\cal PP}}$ is determined from the corresponding two-photon width by replacing the photon coupling $\alpha=e^2/4\pi$ with the pomeron coupling $3\beta_0^2/4\pi$, where 3 is the color factor.  However, such a pointlike quark-pomeron coupling does not account for the necessary decrease in the coupling when one of the quarks is off mass shell.  We thus modulate $\beta_0$ as \cite{DL3}
\begin{eqnarray}
\beta(Q^2) = \frac{\mu_0^2}{\mu_0^2 + Q^2}~\beta_0~,
\end{eqnarray}
where $\mu_0 = 1.1$ GeV, and $Q^2$ measures how far one of the quark legs is off mass shell.
Strictly speaking, $Q$ is an integration variable at the quark-pomeron vertices in Figure 1; we will simply assume $Q=M_X/2$ and use the substitution
\begin{eqnarray}
\beta_0 \rightarrow \tilde{\beta}_0 \equiv \beta(M_X^2/4)
\end{eqnarray}
for the quark-pomeron coupling.
With these choices, the integral in Equation (7) can be performed analytically, giving
\begin{eqnarray}
\sigma^{\cal PP}_{AB} = \left(\frac{3A\beta_0Q_0}{2\pi}\right)^4 
           \frac{9(\tilde{\beta_0}Q_0)^4}{16\pi^2}\frac{1}{\alpha^2}
           \frac{\Gamma_{_{X\rightarrow{\gamma\gamma}}}}{M_X^3}
           \left(\frac{M_X^2s}{m^4}\right)^{2\epsilon} 
           K_0\left(\frac{2M_X^2M^2}{sQ_0^2}\right) ~.
\end{eqnarray}

In calculating the impact parameter dependence of the cross section, we write the total ${\cal PP}$ exchange cross section (7) in the impact parameter representation and fold it with the probability that no inelastic interaction takes place other than double pomeron exchange.  The dependence of the cross section on the impact parameter $b$ is then found by integrating the squared matrix element over all spacetime coordinates save for the transverse distance $b$ between nuclei \cite{us,us2}.  Accounting for the pseudoscalar nature of the $\eta$ mesons as discussed in reference \cite{oxy}, we find
$$
\frac{d\sigma^{{\cal PP}\rightarrow\eta}_{AB}}{d^2b} = 2\pi
  \left(\frac{3A\beta_0}{2\pi^2}\right)^4
  \int\frac{dx_1}{x_1}\frac{dx_2}{x_2}~Q_1^4~Q_2^4~\tilde{Q}^2
   ~e^{-x_1^2M^2/Q_1^2}~e^{-x_2^2M^2/Q_2^2}
  \hspace*{1.5in}  
$$
\begin{eqnarray}
 \hspace*{1in}\times
  \left(\frac{x_1x_2s^2}{m^4}\right)^{2\epsilon}
  \sigma_{_{\cal PP}}^\eta(x_1x_2s)
   ~b^2\tilde{Q}^2~e^{-b^2\tilde{Q}^2/2} 
\end{eqnarray}
with  
\begin{eqnarray}
\tilde{Q}^{-2} = \frac{1}{2}(Q_1^{-2} + Q_2^{-2}) 
\end{eqnarray}
where we introduce the quantities
\begin{eqnarray}
Q_i^{-2} \equiv Q_0^{-2} + 2r_i^2 ~,
\end{eqnarray}
and $r_i$, given by Equation (3), is defined for the invariant subprocess with which the pomeron participates, i.e., $s^\prime = s_1 \approx x_2s$ for the pomeron emitted by nucleus 1, $s^\prime = s_2 \approx x_1s$ for the pomeron emitted by nucleus 2.  
Note that $\sigma_{_{\cal PP}}^\eta$ has dimension (energy)$^{-6}$, which compensates the ``incorrect'' dimensionality of $f_{\cal P}^A(x)$ in Equation (8).  
One of the $x$ integrals in (13) is easily done, courtesy of the $\delta$ function in Equation (9); the other can be performed numerically.

We have calculated the total production cross section $\sigma_{AA}^{{\cal PP}\rightarrow\eta}$ and the differential cross section $d\sigma_{AA}^{{\cal PP}\rightarrow\eta}/d^2b$ for both $\eta_c$ and $\eta_b$ production in the collision of $^{208}$Pb nuclei at both LHC and RHIC energies.  Taking $\Gamma_{\eta_c\rightarrow\gamma\gamma}= 6.3$ keV and $\Gamma_{\eta_b\rightarrow\gamma\gamma}= 0.41$ keV \cite{pdg},
our calculations yield a total cross section of 
$\sigma_{AA}({\cal PP}\rightarrow\eta_c) = 0.76$ mb 
and 
$\sigma_{AA}({\cal PP}\rightarrow\eta_b) = 0.81$ nb at LHC energy;
at RHIC, the numbers are understandably lower: 
$\sigma_{AA}({\cal PP}\rightarrow\eta_c) = 58.3~\mu$b 
and 
$\sigma_{AA}({\cal PP}\rightarrow\eta_b) = 13.4$ pb, respectively.  

These cross sections, of course, are overly optimistic in that they do not account for the effects of inelastic nuclear scattering.  The majority of the inelastic events are expected to occur at small values of the impact parameter $b$;  indeed, the elastic nature of the interaction is maintained only in those collisions in which the two nuclei pass by each other.  Thus it is important to verify that a significant portion of the ${\cal PP}$ cross section extends out to measurably large impact parameters.  As in \cite{oxy}, we have included inelastic scattering effects in two different ways.  One is by applying a geometric cutoff at a minimum impact parameter of $2R$, where $R$ is the nuclear radius ($R\approx 7.1$ fm for $^{208}$Pb).  A more realistic approach accounts for inelastic scattering effects using the Glauber approximation \cite{glauber}.  Our numerical results for both absorption methods are presented in Table 1.  

The dependence of the differential cross sections on impact parameter is compared with our earlier results for $\gamma\gamma$ fusion in Figures 2 and 3, where the dotted curves show the effects of absorption in the Glauber approximation.  For $\eta_c$, we find that although both production modes are of roughly the same order of magnitude at $b=2R$, the pomeron contribution is more strongly suppressed because the probability for an elastic collision becomes large only for $b\geq 16$ fm.  For $\eta_b$ production, however, we find that the smallness of the ${\cal PP}$ cross section largely derives from the reduction $\beta_0\rightarrow\tilde{\beta}_0$ at the quark-pomeron vertex, indicating the difficulty of localizing pomerons within a region of high inverse mass.  Intuitively, this effect makes sense:  the pomeron is an effective description of a virtual two gluon state, and as such has an intrinsic spatial extent on the order of 1 GeV$^{-1}$.  It should be less likely to find such a structure concentrated within the smaller limit $M_{\eta_b}^{-1} = (9.4~$GeV$)^{-1}$.

\acknowledgements
{One of us (AJS) thanks Professor B. M\"uller for useful discussions.  This research was supported by an award from Research Corporation.}

\pagebreak

\section*{Figures}

\newcounter{numb}
\begin{list}%
{FIG. \arabic{numb}: }{\usecounter{numb}\setlength{\rightmargin}{\leftmargin}}

\item The fusion of two pomerons from scattering nuclei.  The effective quark-pomeron coupling, Equation (10), is represented by the circles. 

\item The impact parameter dependence of the differential cross section for $\eta_c$ production in nuclear scattering (a) at the LHC and (b) at RHIC.  The solid curves are for double pomeron exchange and electromagnetic production, as indicated; the dotted curves include absorption effects.  

\item The impact parameter dependence of the differential cross section for $\eta_b$ production in nuclear scattering (a) at the LHC and (b) at RHIC.  The solid curves are for double pomeron exchange and electromagnetic production, as indicated; the dotted curves include absorption effects.

\end{list}

\pagebreak

\section*{Tables}

\newcounter{num}
\begin{list}%
{Table \arabic{num}: }{\usecounter{num}\setlength{\rightmargin}{\leftmargin}}

\item Cross sections for $\eta$-meson production in peripheral collisions of $^{208}$Pb nuclei at RHIC and LHC energies.  $\sigma_{AA}^{tot}$ is the total cross section.  For comparison, inelastic scattering effects have been accounted for in two ways: $\sigma_{AA}(b > 2R)$ is the remaining cross section after applying a cut on impact parameter, whereas $\sigma_{AA}^{el}$ uses the Glauber approximation.
\end{list}

\pagebreak

\begin{center}
\begin{tabular}{|c|c|ll|ll|}\cline{3-6}
\multicolumn{2}{c}{}
 &\multicolumn{2}{|c|}{$\eta_c$} &\multicolumn{2}{|c|}{$\eta_b$} \\ \hline
~3.5 TeV/u~& ~~$\sigma_{AA}^{tot}$~  &~0.76& mb~  &~0.81& nb~\\ \cline{2-6}
  ~(LHC) ~ &~$\sigma_{AA}(b>2R)$~~   &~3.4& nb~  &~3.7& fb~\\ \cline{2-6}
           &   $\sigma_{AA}^{el}$    &~8.6& $\mu$b~  &~0.93& pb~\\ \hline
~100 GeV/u~&~~$\sigma_{AA}^{tot}$~  &~58.3&$\mu$b &~13.4& pb~    \\ \cline{2-6}
   ~(RHIC)~&~$\sigma_{AA}(b>2R)$~~   &~0.79& nb &~0.2& fb~    \\ \cline{2-6}
           &  ~$\sigma_{AA}^{el}$~   &~62.4& nb &~14.7& fb~    \\ \hline
\end{tabular}

\vspace*{2in}
Table 1
\end{center}

\end{document}